\documentclass[
    ,final            % use final for the camera ready runs
,aps,nofootinbib,amsmath,floatfix  ]
  {aipproc}

\layoutstyle{6x9}

\usepackage{amsmath}
\usepackage{amssymb}  % \gtrsim, \geqslant, etc etc: see amsguide.ps
\usepackage{graphicx}
\usepackage[dvips]{color}  % Only for markup of new and old text

\newcommand{\beq}{\begin{equation}}
\newcommand{\eeq}{\end{equation}}
\newcommand{\ba}{\begin{array}}
\newcommand{\ea}{\end{array}}
\newcommand{\bea}{\begin{eqnarray}}
\newcommand{\eea}{\end{eqnarray}}
\newcommand{\bi}{\begin{itemize}}  %\setlength{\itemsep}{0\parsep}}
\newcommand{\ei}{\end{itemize}}
\newcommand{\ben}{\begin{enumerate}} %\setlength{\itemsep}{0\parsep}}
\newcommand{\een}{\end{enumerate}}
\newcommand{\bc}{\begin{center}}
\newcommand{\ec}{\end{center}}
\newcommand{\non}{\nonumber\\}

\usepackage[normalem]{ulem}  % \sout{old text} for strikeout

\begin{document}

\title{Bulk viscosity in 2SC and CFL quark matter}


\author{Mark G.\ Alford }{
  address={Department of Physics, Washington University St Louis, MO, 63130, USA}
}

\author{Andreas Schmitt}{
  address={Department of Physics, Washington University St Louis, MO, 63130, USA}
}

\begin{abstract}
The bulk viscosities of two color-superconducting phases, the color-flavor locked (CFL) phase and the 2SC
phase, are computed and compared to the result for unpaired quark matter. In the case of the CFL phase, 
processes involving kaons and the superfluid mode give the largest contribution to the bulk viscosity since
all fermionic modes are gapped. In the case of the 2SC phase, ungapped fermionic modes are present and 
the process $u+d\leftrightarrow u+s$ 
provides the dominant contribution. In both cases, the bulk viscosity can become larger than that of the 
unpaired phase for sufficiently large temperatures ($T\gtrsim 1$ MeV for CFL, $T\gtrsim 0.1$ MeV for 2SC). 
Bulk viscosity (as well as shear viscosity) 
is important for the damping of $r$-modes in compact stars and thus can potentially be used as an indirect 
signal for the presence or absence of color-superconducting quark matter. 
\end{abstract}

\maketitle

%%%%%%%%%%%%%%%%%%%%%%%%%%%%%%%%%%%%%%%%%%%%
%% MAINMATTER
%%%%%%%%%%%%%%%%%%%%%%%%%%%%%%%%%%%%%%%%%%%%

{\em Color superconductivity in compact stars. --}
Matter at sufficiently large densities and low temperatures is a color superconductor, which 
is a degenerate Fermi gas of quarks with a condensate of Cooper pairs near the Fermi surface
\cite{Bailin:1983bm}. At asymptotically large densities,
where the quark masses are negligibly small compared to the quark chemical potential $\mu$, 
three-flavor quark matter is in the color-flavor locked (CFL) state \cite{Alford:1998mk}. In this 
state quarks form Cooper pairs in a particularly symmetric fashion. The gauge group of the strong 
interactions $SU(3)_c$ and the chiral symmetry group $SU(3)_L\times SU(3)_R$ are spontaneously broken down
to the group of joint color-flavor rotations $SU(3)_{c+L+R}$. All quarks participate in pairing, giving 
rise to energy gaps in all quasifermion modes. Therefore, the lightest degrees of freedom in the 
CFL phase are 
Goldstone bosons: the broken chiral symmetry gives rise to an octet, with the lightest modes
being the charged and neutral kaons (at not asymptotically large densities, chiral symmetry is 
an approximate
symmetry and the Goldstone modes acquire a small mass). Moreover, the CFL phase is a superfluid, spontaneously
breaking baryon number conservation symmetry $U(1)_B$. Hence there is a ``superfluid mode'' which 
is exactly massless. Goldstone modes in the CFL phase can be described within effective 
theories \cite{Casalbuoni:1999wu,Son:1999cm,Son:2002zn}, not unlike conventional chiral perturbation theory of
the QCD vacuum. The results given below for the CFL bulk viscosity are computed within these effective 
theories. 

While rigorous QCD calculations from first principles can be done at asymptotically large densities,
the situation is more complicated at moderate densities. Here we are interested in densities of matter inside
a compact star. These densities can be as large as several times nuclear ground state density; 
however, even then 
the quark chemical potential is at most of the order of 500 MeV. Therefore, perturbative methods within 
QCD are not applicable and
one has to rely on more phenomenological models. Furthermore, the mass of the strange quark $M_s$ is not
negligible. This fact, together with the conditions of electric and color neutrality, 
imposes a ``stress'' on the CFL phase. By stress we mean here that the Fermi momenta of the quarks that 
pair are no longer equal. A sufficiently large stress (i.e., a sufficiently large mismatch in Fermi momenta 
compared to the pairing gap) will eventually 
disfavor the CFL phase; however, the density at which this happens is unknown. Unless
the CFL phase is directly superseded by nuclear matter, less symmetric color-superconducting phases  
are expected to form the ground state. One of these phases is the so-called 2SC phase 
\cite{Bailin:1983bm}. In the 2SC phase, all strange quarks as well as up and down quarks of one color 
remain unpaired. This is one possible way for the system to deal with the heaviness of the strange quark. 
In phase diagrams based on Nambu-Jona-Lasinio models, the 2SC phase appears to be a candidate for the
ground state at moderate densities \cite{Ruster:2005jc,Abuki:2005ms}.
Here we shall not discuss other interesting possibilities such as Larkin-Ovchinnikov-Fulde-Ferrell 
(LOFF) states \cite{Alford:2000ze,Rajagopal:2006dp} or spin-triplet pairing 
\cite{Schmitt:2004et,Schafer:2000tw} which break translational and/or rotational invariance.
For a calculation of the bulk viscosity in spin-triplet phases, see Ref.\ \cite{Sa'd:2006qv}.
 
The only system in nature where deconfined quark matter may be expected, and hence the only 
``laboratory'' for the study of color superconductivity, is the interior of compact stars. The
temperature of compact stars drops below $\sim 1$ MeV within the first minute of the star's life.
Hence, quark matter, if present in the core of the star, 
can be expected to be in a color-superconducting state,
since the critical temperature for spin-singlet color superconductors, such as the 2SC and CFL phases, 
is estimated to be of the order of tens of MeV. 
Consequently, it is of great interest to compute thermodynamical and transport properties of 
different candidate color 
superconductors.
Results of these calculations can be used to predict the behavior of the star with a quark matter core 
in the respective phase. Comparing these predictions with the actual astrophysical observations 
then may serve to rule out certain possibilities for the ground state of quark matter and thus eventually
lead to a better understanding of the phases of QCD at high densities. 
One observable which is sensitive to the 
micsoscopic properties of the matter inside the star is for example 
the temperature as a function of time, i.e., 
the cooling curve of a star. Since the dominant cooling mechanism within the first million years in the 
evolution of a compact star is provided by energy loss
through neutrinos, calculations of the neutrino emissivity (and the specific heat) have been performed
for a variety of color superconductors, see for instance 
\cite{Schmitt:2005wg,Jaikumar:2005hy,Jaikumar:2002vg}. The remainder of these proceedings is devoted to 
another transport property, namely bulk viscosity, which is related to the rotation frequency of the 
star.   

\bigskip
{\em Bulk viscosity and $r$-modes. --} A compact star exhibits vibrational or rotational modes, 
typically in the kHz range \cite{Kokkotas:2001ze}.
The spectrum of these pulsations is very rich. They have been classified according to their restoring
force. For instance, so-called $r$-modes are associated with the coriolis force. These modes are of particular
interest because they are unstable with respect to emission of gravitational waves 
for all perfect fluid stars \cite{Andersson:1997xt,Andersson:2001bz}. For a nice pedagogical 
introduction on neutron star pulsations and instabilities, see Ref.\ \cite{Lindblom:2000jw}.
If a neutron star was a perfect fluid, $r$-mode instabilities would lead to 
a drastic and quick spin-down, since the rotational energy would be transformed into the 
energy of the gravitational waves. The fact that fast rotating stars are observed leads to the 
conclusion that there must be a damping mechanism for these instabilities. This damping
can be provided by the viscosity of the fluid. 

\begin{figure}[ht]
\includegraphics[width=0.6\hsize]{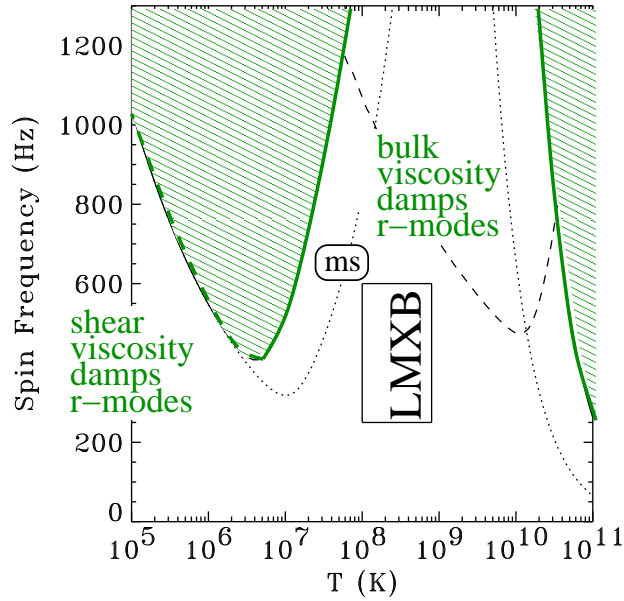}
\vspace{-0.5cm}\caption{Damping effects of shear and bulk viscosity of unpaired quark matter in the rotation 
frequency -- temperature diagram for a strange mass $M_s=200$ MeV. Shaded areas are forbidden due to $r$-mode 
instabilities. The damping effects are dominated by shear viscosity for temperatures lower than $T\sim 10^6$K,
while they are dominated by bulk viscosity for larger temperatures, $10^7\,{\rm K}< T< 
10^{10}\,{\rm K}$. The dotted lines correspond to a different strange mass, $M_s=100$ MeV. The dashed line
in the middle of the diagram corresponds to nuclear matter. ms = millisecond pulsars, 
LMXB = low-mass X-ray binaries. The figure is taken (and slightly modified) from Ref.\ \cite{Madsen:1999ci}.}
\label{fig:rmode}
\end{figure}

Both shear and bulk viscosity can provide damping. In Fig.\ \ref{fig:rmode} we show that they 
are important in different temperature regimes in the example of unpaired
quark matter. In the following, we discuss if and how color superconductivity may change this picture.  
Calculations of bulk viscosity exist for conventional neutron star matter
\cite{Haensel:2001mw,Haensel:2000vz,Sawyer:1989dp}, 
hyperonic matter \cite{Haensel:2001em,Lindblom:2001hd}, and kaon condensed hadronic matter
\cite{Chatterjee:2007qs}. For bulk viscosity in unpaired quark matter, see 
Refs.\ \cite{Anand:1999bj,Madsen:1992sx,Dong:2007mb,Dong:2007ax,Sa'd:2007ud}. 

\bigskip
{\em Definition of bulk viscosity. --}
The bulk viscosity $\zeta$ is defined as
\beq \label{defbulk}
\zeta \equiv 2\left\langle\frac{dW}{dt}
\right\rangle\left(\frac{V_0}{\delta V_0}\right)^2\frac{1}{\omega^2} \, .
\eeq
It is proportional to the average dissipated power per volume $\langle dW/dt\rangle$ 
in one oscillation period $\tau = 2\pi/\omega$.
In the present case, (local) volume oscillations are created by $r$-modes, whose typical
frequencies are of the order of the rotation frequency of the star, $\omega/(2\pi)\lesssim 1{\rm ms}^{-1}$.
(Another application of bulk viscosity is radial pulsations of neutron stars 
\cite{Sahu:2001iv,Haensel:2002qw}.) We denote the volume oscillation by $V(t)=V_0 + \delta V_0\cos\omega t$.
In general, a superfluid has more than one bulk viscosity coefficient, since there are contributions 
related to the stresses in the superfluid flow relative to the normal one 
\cite{Gusakov:2007px,Andersson:2006nr,khala}. We shall neglect these effects and compute the single 
bulk viscosity given in Eq.\ (\ref{defbulk}). (In the usual terminology of superfluid hydrodynamics
our $\zeta$ is equivalent to the coefficient $\zeta_2$.)

Let us first consider a system where the 
lightest degrees of freedom are fermions, e.g., unpaired and 2SC quark matter. 
The volume oscillation forces the system out of chemical equilibrium with respect to the nonleptonic process
\beq \label{udus}
u + d \leftrightarrow u + s
\eeq
and the semi-leptonic processes
\begin{subequations} \label{semilept}
\bea
u + e &\leftrightarrow& d + \nu_e \\
u + e &\leftrightarrow& s + \nu_e \, .
\eea
\end{subequations}
In other words, a volume oscillation causes oscillations in the quantities $\delta\mu_1\equiv\mu_s-\mu_d$,
$\delta\mu_2\equiv\mu_d-\mu_u-\mu_e$, $\delta\mu_3\equiv\mu_s-\mu_u-\mu_e$ (only two of 
which are independent). 
In the following, we ignore the semi-leptonic processes, which are in general slower than the nonleptonic
one, and denote $\delta \mu\equiv \delta\mu_1$. The effect of the semi-leptonic processes has been 
discussed in Refs.\ \cite{Sa'd:2007ud,Alford:2006gy}. 
The weak interaction provides the dominating processes since
they ``resonate'' best with the external oscillation frequency. In principle, strong interactions 
also provide a contribution to the bulk viscosity, e.g., through thermal reequilibration. However, the
speed of these processes compared to the time scale set by the oscillation renders this
contribution negligible. 

The dissipated power is related to the oscillations in volume and chemical potential,
\beq \label{resultpower}
\left\langle\frac{dW}{dt}\right\rangle = B\lambda\,\langle\delta\mu(t)\,\delta v(t)\rangle
\, .
\eeq
The coefficient $\lambda$ has to be determined from the 
microscopic calculation. It is defined by $\Gamma_d[\delta\mu(t)] = \lambda\, \delta\mu(t)$, 
where $\Gamma_d[\delta\mu(t)]$ is the number of $d$ quarks per volume and time produced in the reaction
$u+s\leftrightarrow u+d$. The linear dependence of $\Gamma_d$ on $\delta\mu$ is 
a valid approximation for sufficiently small volume oscillations,  
$\delta V_0 \ll V_0$. The coefficient $B$ is related to the changes of the equilibrium 
particle densities with respect
to the corresponding chemical potentials and depends on the phase. We have abbreviated 
$\delta v\equiv \delta V_0/V_0\,\cos\omega t$.
Finally, the relation between $\delta\mu$ and $\delta v$ is given by the differential equation
\beq \label{diffeq}
\frac{\partial \delta\mu}{\partial t}
= B\,\frac{\partial \delta v}{\partial t} - \gamma \,\delta\mu(t)  \, , 
\eeq
where $\gamma\equiv C\lambda$  is the characteristic inverse
time scale of the microscopic process. The coefficient $C$ depends, as the coefficient $B$ above, 
only on the equilibrium densities and has to be determined for each phase 
separately. As pointed out in Ref.\ \cite{Alford:2006gy}, 
Eqs.\ (\ref{resultpower}) and (\ref{diffeq}) show that the present system is completely analogous to 
an electric circuit with alternating voltage -- playing the role of  
$\delta v(t)$ -- and alternating current
 -- playing the role of $\delta\mu(t)$. In this
analogy, resistance $R$ and capacitance $C'$ can be identified with microscopic 
quantities of the quark system, $R\to 1/B$, $C'\to B/\gamma$. 

Solving the differential equation (\ref{diffeq}) and inserting the result into Eq.\ 
(\ref{resultpower}) and then 
into (\ref{defbulk}) yields
\beq \label{bulkfinal}
\zeta = \frac{B^2}{C}\frac{\gamma}{\gamma^2+\omega^2} \, . 
\eeq
This expression for the bulk viscosity shows that, for a given $\omega$, $\zeta$ has a maximum at
$\gamma = \omega$. Consequently, bulk viscosity is a resonance phenomenon with the 
largest bulk viscosity occurring when the time scale of the
microscopic process matches the one of the external oscillation. 

In the CFL phase, the contributions to the bulk viscosity from the processes (\ref{udus}) and (\ref{semilept})
are exponentially suppressed and thus do not play any significant role for the bulk viscosity at any 
temperature \cite{Madsen:1999ci}. The dominant contribution rather comes from the lightest Goldstone 
modes, the 
superfluid mode $\varphi$, originating from breaking baryon number conservation symmetry, and 
the neutral kaon $K^0$, originating from the breaking of chiral symmetry. 
The contribution to $\zeta$ from the process 
\beq \label{HHH}
\varphi \leftrightarrow \varphi + \varphi
\eeq
has been computed in Ref.\ \cite{Manuel:2007pz}. While the $\varphi$ is 
exactly massless, the mass of the $K^0$ can be deduced from the effective theory. The result, however, 
depends on the quark masses, the fermionic energy gap, and the quark chemical potential and is, 
strictly speaking, only valid at asymptotically large densities. Therefore, the (zero-temperature)
kaon mass $m_{K^0}$ (and the 
effective kaon chemical potential $\mu_{K^0}$) are poorly known. In particular, it is not known 
whether the kaon mass and chemical potential allow for Bose-Einstein condensation. 
Condensation occurs for $\mu_{K^0}>m_{K^0}$, which seems likely from extrapolating the high density 
results to densities relevant for compact stars. If indeed the inequality  
$\mu_{K^0}>m_{K^0}$ is true, it can be expected that a kaon condensate is present for all 
relevant temperatures, since
the critical temperature for condensation within the effective theory has been estimated to be of the 
order of tens of MeV \cite{Alford:2007qa}. In the following, however, 
we shall assume that $\mu_{K^0}<m_{K^0}$, i.e., there are only thermal kaons in the system. The bulk 
viscosity in the presence of a condensate has not yet been computed. 

The contribution of the process
\beq \label{KHH}
K^0\leftrightarrow \varphi + \varphi 
\eeq
has been studied in Ref.\ \cite{Alford:2007rw}. This process reequilibrates the system with respect to the 
``out-of-equilibrium'' quantity $\delta\mu_{K^0}\equiv \mu_d - \mu_s$, since $K^0$ carries
quantum numbers $\bar{s}d$.

\begin{figure}[t]
\includegraphics[width=\hsize]{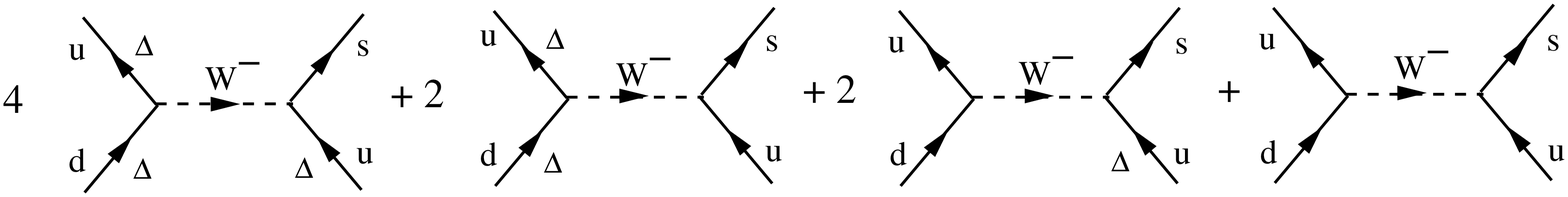}
\caption{Contributions to the process $u+d\to u+s$ in the 2SC phase. A gapped fermion is marked with the
gap $\Delta$ at the respective line. We have omitted (small) contributions from anomalous propagators.}
\label{fig:diagrams}
\end{figure}

\bigskip
{\em Rate of nonleptonic processes. --}
Before we present the results for the bulk viscosity for the 2SC and CFL phases, we 
briefly outline the calculation of the rate of the nonleptonic process (\ref{udus}).  
This calculation is done with a fixed nonzero $\delta\mu$, yielding different rates for the two directions 
$u+d\to u+s$ and $u+s\to u+d$. Hence one obtains a net production rate $\Gamma_d$ of $d$ quarks. 
In the 2SC phase, $\Gamma_d$ is computed from the diagrams (and their reverse directions)
shown in Fig.\ \ref{fig:diagrams}. The combinatorical factors 
in front of the diagrams are obtained upon counting color degrees of freedom: one can 
attach one of three colors to each of the two weak vertices, giving rise to 9 possibilities.
In the 2SC phase all blue quarks and all strange 
quarks are unpaired while all other modes are paired. Consequently, 4 of the 9
possibilities contain three gapped modes (red or green for both vertices), 2 contain two gapped modes 
(red or green for one, blue for the other vertex), 2 contain one gapped
mode (blue for one, red or green for the other vertex), and one contains only unpaired modes (blue for both 
vertices). 
Therefore, at temperatures much smaller than the fermionic energy gap, $T\ll\Delta$,   
where the contributions of gapped quarks are exponentially 
suppressed, $\Gamma_d$ is to a very good approximation given by \cite{Madsen:1999ci}
\beq 
\label{factor9}
\Gamma_d^{\rm 2SC} = \frac{1}{9}\Gamma_d^{\rm unp} \qquad 
   {\rm for} \;\; T\ll \Delta \, , 
\eeq
since only the one reaction containing only unpaired modes
contributes. The rate $\Gamma^{\rm unp}_d$ was computed in \cite{Madsen:1993xx}.
For larger temperatures, the contributions from the gapped modes cannot be neglected. 
They become important for temperatures
$0.3\,T_c\lesssim T<T_c$. Each diagram yields a contribution which, 
for one direction of the process, schematically reads (for the exact expression and its derivation see
Ref.\ \cite{Alford:2006gy})
\bea
\Gamma_d &\propto&\sum_{\{e_i\}}\int_{\{p_i\}}{\cal F}\,\delta(e_1\epsilon_1
+e_2\epsilon_2-e_3\epsilon_3-e_4\epsilon_4+\delta\mu) \non
&&\times\; f(e_1\epsilon_1)f(e_2\epsilon_2) f(-e_3\epsilon_3) f(-e_4\epsilon_4) \, .
\eea 
Here, ${\cal F}$ is a complicated function of the momenta $p_i$  and 
the signs $e_i=\pm 1$, $i\le 4$; $\epsilon_i$ are
the quasiparticle energies, and $f$ the Fermi distribution functions. The sum over the signs $e_i$
is typical for collision integrals for any superfluid or superconductor (see for instance 
Ref.\ \cite{PhysRevB.15.3367}
in the context of superfluid $^3$He): 
the process $u+d\to u+s$ does not only get contributions 
from two quasiparticles $u$, $d$, going into two quasiparticles $u$, $s$. There are also contributions from 
three particles coalescing into one particle (and one particle decaying into three) and from 
annihilating (and creating) four particles. This is possible because quasiparticles can be created from 
or deposited into the condensate. Formally, this is obvious from the Fermi distribution 
functions where the factor $f(\epsilon)$ accounts for an ingoing particle and 
$f(-\epsilon)=1-f(\epsilon)$ for an outgoing particle. Due to the sum over $\{e_i\}$, all combinations
of the $f$'s can appear, e.g., $f(\epsilon_1)f(\epsilon_2)f(\epsilon_3)[1-f(\epsilon_4)]$.
Here we shall not present the explicit results for the rate itself but rather directly skip to 
the result for the bulk viscosity, which is a function of this rate.

\begin{figure}[ht]
\includegraphics[width=0.65\hsize]{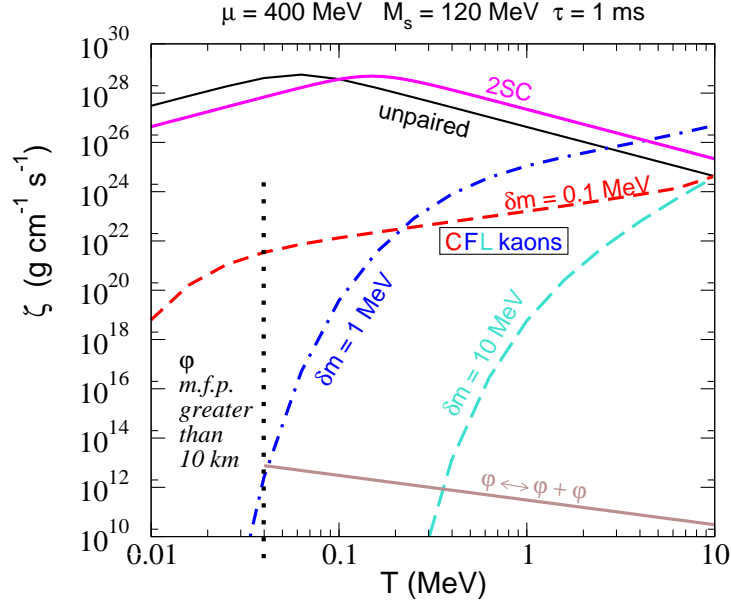}
\caption{Bulk viscosities as functions of temperature for an oscillation period $\tau=2\pi/\omega=1$ ms.
CFL phase: contribution
from the process $K^0\leftrightarrow \varphi+\varphi$ for different values of 
$\delta m\equiv m_{K^0}-\mu_{K^0}$ and contribution 
from $\varphi\leftrightarrow\varphi+\varphi$, see Eq.\ (\ref{zeta_cfl}). 
2SC phase and unpaired quark matter:
contribution from the process $u+d\leftrightarrow u+s$.}
\label{fig:bulkCFL}
\end{figure}

\bigskip
{\em Results for 2SC and CFL quark matter. --}
Inserting the result for the net production rate of $d$ quarks into (\ref{bulkfinal}) yields the
bulk viscosity for the 2SC phase as a function of temperature. 
The result for a typical oscillation frequency $\tau\equiv 2\pi/\omega = 1$ ms is shown 
in Fig.\ \ref{fig:bulkCFL}. A critical temperature
of $T_c=30$MeV is assumed. For low temperatures, 
the time scale of the nonleptonic process is much smaller than the oscillation frequency
$\gamma\ll\omega$, implying $\zeta\propto \gamma$. Consequently, from Eq.\ (\ref{factor9})
we conclude $\zeta_{\rm 2SC}=\zeta_{\rm unp}/9$. For large temperatures, however, we have     
$\gamma\gg\omega$ and thus $\zeta\propto 1/\gamma$. Consequently, the superconducting phase, which 
has the slower rate, has the larger bulk viscosity. The result becomes nontrivial (= $\zeta_{\rm 2SC}$
not being simply a numerical factor times $\zeta_{\rm unp}$) between these two regimes, i.e., 
$\gamma\sim \omega$, and for temperatures close
to the critical temperatures, not shown in this figure (see Ref.\ \cite{Alford:2006gy}). 

The bulk viscosity from the process (\ref{KHH}) in the CFL phase is obtained along the same lines. 
One inserts the result for the rate of the kaonic process (\ref{KHH}) into Eq.\ (\ref{bulkfinal}). 
The result is shown in Fig.\ \ref{fig:bulkCFL} for different values of $\delta m\equiv m_{K^0}-\mu_{K^0}$
(remember that we assume that there is no kaon condensate, i.e., $\delta m >0$). We have also plotted
the contribution from the process (\ref{HHH}), which, in the limit $\omega\to 0$ is given by 
\cite{Manuel:2007pz}
\beq \label{zeta_cfl}
\zeta^{\varphi}_{\rm CFL} = 0.011\frac{M_s^4}{T} \, . 
\eeq

We observe that 
at $T\simeq (1-10)$ MeV the bulk viscosity of CFL matter is 
comparable to that of unpaired quark matter. For $T<1$~MeV, $\zeta$
is strongly suppressed. Depending on the poorly known value for $\delta m$
the pure $\varphi$ contribution (\ref{zeta_cfl}) may dominate over the contribution from the 
$K^0\leftrightarrow \varphi+\varphi$ reaction at low enough temperatures. 
However, for $T<0.1$ MeV the $\varphi$ mean free 
path is on the order of the size of the star, i.e., the system 
is in the collisionless rather than in the hydrodynamic regime, and the 
result ceases to be meaningful.

\begin{theacknowledgments}
This research was
supported by the U.S.~Department of Energy under contracts
\#DE-FG02-91ER40628,  % Wash U theory
\#DE-FG02-05ER41375 (OJI). % Mark's OJI

\end{theacknowledgments}



\bibliographystyle{aipproc}   % if natbib is available

\bibliography{schmitt}

\end{document}